\documentclass [11pt]{article}
\usepackage[dvips]{graphicx}
\newtheorem{property}{Property}
\newcommand{\Tr}{\mathop{\mathrm{Tr}}\nolimits}

\begin{document}
\title {\bf Scalar charged particle \mbox{in Weyl--Wigner--Moyal phase
space.}\\ Constant magnetic field\footnote{{\it Journal of Russian
Laser Research} (Kluwer Academic/Plenum Publishers) {\bf 23}, no
4, P. 347--368 (2002)}}
\author{B.I. Lev$^{1,2}$\footnote{E-mail: lev@iop.kiev.ua},
A.A. Semenov$^{1}$\footnote{E-mail: sem@iop.kiev.ua}, C.V.
Usenko$^{2}$\footnote{E-mail: usenko@ups.kiev.ua}
\\
\parbox[c]{0.95\textwidth} {\center\textsl{\small $^{1}$
Institute of Physics, National Academy of Sciences of Ukraine,\\
46 Nauky pr, Kiev 03028, Ukraine}} \\
\parbox[c]{0.95\textwidth}{\center\textsl{\small $^{2}$ Physics Department, Taras
Shevchenko Kiev University,\\ 6 Academician
 Glushkov pr, Kiev 03127, Ukraine}}}
\date{}
\maketitle

\begin{abstract} A relativistic phase-space representation
for a class of observables with matrix-valued Weyl symbols
proportional to the identity matrix (charge-invariant observables)
is proposed. We take into account the nontrivial charge structure
of the position and momentum operators. The evolution equation
coincides with its analog in relativistic quantum mechanics with
nonlocal Hamiltonian under conditions where particle-pair creation
does not take place (free particle and constant magnetic field).
The differences in the equations are connected with peculiarities
of the constraints on the initial conditions. An effective
increase in coherence between eigenstates of the Hamiltonian is
found and possibilities of its experimental observation are
discussed.
\end{abstract}

\section{Introduction}
\label{s1}

In this paper, we continue our previous study~\cite{b1} of the
phase-space representation for relativistic quantum mechanics. We
have established two problems in developing the
Weyl--Wigner--Moyal (WWM) formalism for the relativistic case.
First of all, it is worth noting that  Weyl transformation does
not include time as a dynamical variable, i.e., it is not Lorentz
invariant. The second problem relates to the existence of a charge
variable, which is a specific degree of freedom in relativistic
quantum mechanics~\cite{b2}.  This degree of freedom appears due
to the
 procedure of canonical quantization of the relativistic
particle~\cite{b2a}; the presence of this degree of freedom
results in the fact that the standard position operator is
 not well
defined~\cite{b2,b3,b4,b4a}. Therefore, we points out the
nontrivial charge structure of the position operator  here. In the
context of the phase-space quantization, this problem is still
open.

{\hfuzz=4,5pt Nowadays, different approaches to solve these
problems are well known. Each of them has at least two alternative
solutions with different interpretations. In short, we emphasize
approaches where the problem of Lorentz invariance is solved by
the generalization of Weyl transformation over the whole
space--time (stochastic interpretation of quantum mechanics, de
Groot--van Leeuwen--van Weert's approach \cite{b5}), and those
ones where Weyl transformation is applied in the three-dimensional
space or on a space-like hyper-surface. The second problem can be
solved in two ways --- using either the standard position operator
or Newton--Wigner position operator~\cite{b3}. A list of
references on the problem can be found in our previous work
\cite{b1}; here we would like to complete the list.

}The relativistic Wigner function for spin-1/2 particles in
magnetic field was introduced in \cite{b6}. This approach was
realized using the standard (three-dimensional) Weyl
transformation and Dirac equation; therefore, one can say, that
the standard position operator was used.

The consideration based on Weyl transformation in the
three-dimen\-sional space was considered in \cite{b7} for the
semirelativistic approximation.

The Wigner function for the covariant harmonic oscillator and for
the light waves was treated in \cite{b8}; while in \cite{b8aa} it
was shown, that space--time geometry of relativistic particles
corresponds to the four-dimensional phase space of the coupled
oscillators, and the corresponding Wigner function was presented.

In our consideration \cite{b1}
 of a spin-0 free particle,
we used the following assumptions:
\begin{enumerate}
\item\label{i1} Weyl transformation in the
six-dimensional phase space (related to the three-dimensional
configuration space) or, more generally, on the spacelike
hypersurface in the context of Tomonaga--Schwinger app\-roach to
quantum field theory \cite{b9}, is considered;
\item\label{i2} The standard position operator
instead of Newton--Wigner position operator is used.
\end{enumerate}

The argumentation in favor of (\ref{i1}) follows. The mean values
calculated in this approach coincide with those calculated in the
usual (Schr\"odinger) representation, which is Lorentz invariant.
The three-dimensional integration is a consequence of the fact,
that the scalar product of states in the Schr\"odinger
representation is determined on a space-like hyper-surface. It
cannot be redefined in the whole space--time domain without
additional physical assumptions. Furthermore, it is worth noting
that this approach can be explained within the framework of the
concept of a reference frame in which a reduction of the quantum
state takes place (for a details and review of recent experiments
see~\cite{b8a}).

The argumentation in favor of (\ref{i2}) is the fact that the
approach using the Newton--Wigner position operator is not Lorentz
invariant even in the Schr\"odinger representation. Here one faces
more difficult problems concerning the definition of mean values
and the wave equation itself. On the other hand, using the
standard position operator enables one to define the probability
density for localized states in the domain of the positive
(negative) energy \cite{b8b}.

We have considered spin-0 particles using the Feshbach--Villars
formalism \cite{b2} and restricted ourselves by the class of observables
for which their matrix-valued Weyl symbols are proportional to the identity
matrix (charge-invariant observables).

The aim of the paper is to generalize this approach to a particle
in a static magnetic field. This case, like the free particle
case, admits of the one-particle interpretation of relativistic
quantum mechanics. We pay special attention to the problem of the
influence of the nontrivial charge structure of the position (and
momentum) operator on the mean values of the charge-invariant
observables.

Contrary to the free particle case, not only the position operator
but the momentum operator as well have the nontrivial charge
structure~\cite{b10}. The generalization of the Newton--Wigner
position operator approach is the theory with a nonlocal
Hamiltonian~\cite{b11}
\begin{equation}
\hat{E}=\sqrt{m^2c^4+c^2\left[\hat{p}-eA\left(\hat{q}\right)\right]^2};
\label{f1}\end{equation} here we call it the nonlocal
 theory. It is worth noting, that term
``nonlocal'' has several meanings in modern quantum physics. We
use it here for
 pointing out the fact, that differential equations contain
derivatives up to infinite order.

Contrary to the case of nonstationary electromagnetic field, the
Hamiltonian (\ref{f1}) can be redefined by extending its
action on the charge domain of the Hilbert space
\begin{equation}
\hat{H}^{nl}=\tau_3\hat{E} \label{f2}\end{equation}
 and it can be considered as
 the Hamiltonian of the standard theory
\begin{equation}
\hat{H}=\left(\tau_3+i\tau_2\right)\frac{\left(\hat{p}
-eA\left(\hat{q}\right)\right)^2}{2m}
+\tau_3mc^2 \label{f3}
\end{equation}
in a particular representation that we call the nonlocal theory
representation.
 Here and bellow $\tau_i$ are the 2$\times$2 matrices defined in \cite{b2}.
 Generally speaking, the operators $\hat{p}$ and $\hat{q}$
from (\ref{f1}) differ from the standard momentum and position operators,
 which are employed in (\ref{f3}).
Therefore, we call $\hat{p}$ and $\hat{q}$ from (\ref{f1}) the
momentum and position operators of nonlocal theory. Contrary to
the free particle case, these operators are not even parts of the
standard position and momentum operators. If
$A\left(\hat{q}\right)=0$, i.e., in the absence of magnetic field,
expression (\ref{f2}) is the Hamiltonian of free particle in the
 Newton--Wigner position operator representation.

In Section \ref{s2} we consider the general properties of algebra
of charge-invariant observables. In Section \ref{s3} we introduce
the four-component (related to a particle, anti-particle and two
interference components) Wigner function for
 charge-invariant observables and consider its
evolution equation. The Heisenberg picture of motion and an
extension of the concept of  Weyl symbol for charge-invariant
observables, which in our approach contains four components, are
given in Section \ref{s4}. Section \ref{s5} is devoted to the
consideration of specific constraints on the Wigner function for
charge-invariant observables, which is a basic difference between
the standard and nonlocal theories. This peculiarity has a simple
physical sense and is responsible for an effective increase in
coherence between eigenstates of the Hamiltonian; this is
considered in Section \ref{s6} along with simple examples
 and a proposal of the possibility of
experimental observation.

\section{Charge-invariant observables}
\label{s2}

For a consistent development of the WWM formalism, one should
define Weyl transformation. Following \cite{b1} we describe an
operator $\hat{A}$ by the 2$\times$2 operator-valued matrix
$\hat{A}_\alpha{}^\beta$, and the corresponding matrix-valued Weyl
symbol by $c$-number matrix $A_\alpha{}^\beta(p,q)$. The Greek
indices take values $\pm1$ and, whenever possible, we will label
them as $\pm$. Since the Hilbert space of states for scalar
charged particles has an indefinite metric, one should distinguish
between covariant and contravariant indices. Taking the above
mentioned into account, one can write the Weyl transformation as
follows
 \begin{equation}
\hat{A}_\alpha{}^\beta=\sum_{\gamma=\pm1}\int
\limits_{-\infty}^{+\infty}A_\gamma{}^\beta\left(p,q\right)
\hat{W}_\alpha{}^\gamma\left(p,q\right)dp\,dq, \label{f4}
\end{equation}
where $\hat{W}_\alpha{}^\beta\left(p,q\right)$ is the operator of
quasiprobability density.

Let us introduce the momentum (or position) part of the
eigenfunction of the Hamiltonian (\ref{f3}) as the solution of the
following eigenvalue problem:
\begin{equation}
\frac{1}{2m}\left(\hat{p}-
eA\left(\hat{q}\right)\right)^2\varphi_n\left(p\right)
=e(n)\varphi_n\left(p\right), \label{f5}
\end{equation}
where $n$ is a set of quantum numbers. The eigenvalue modulus of
Hamiltonians (\ref{f1})--(\ref{f3}) can be expressed
through $e(n)$
\begin{equation}
E(n)=mc^2\sqrt{1+\frac{2}{mc^2}e(n)}. \label{f6}
\end{equation}

In the representation of these functions, Weyl transformation
(\ref{f4}) has the following form:
\begin{equation}
A_{nm;\alpha}{}^\beta=\int\limits_{-\infty}^{+\infty}A_\alpha{}^\beta
\left(p,q\right)
W_{nm}\left(p,q\right)dp\,dq \label{f7}
\end{equation}
where $W_{nm}\left(p,q\right)$ is the Hermitian generalization of
the Wigner function \cite{b12,b13}
\begin{equation}
W_{nm} (p,q) = \frac{1}{{(2\pi \hbar )^d }}\int\limits_{ - \infty
}^{ + \infty } {\varphi _m^ *  \left( {p + \frac{P}{2}}
\right)\varphi _n \left( {p - \frac{P}{2}} \right)\exp\left( -
\frac{i}{\hbar }Pq\right) dP}\label{f8}
\end{equation}
with $d$ being the dimensionality of the configuration space. The
final transformation to the energy representation, where the
matrix of the Hamiltonian (\ref{f3}) has a diagonal form, is
realized by the following transformation matrix:
\begin{equation}
U_{nm;\alpha}{}^\beta=\frac{1}{2\sqrt{mc^2E(n)}}\left[\left(E(n)+mc^2\right)
\delta_\alpha{}^\beta+\left(E(n)-mc^2\right)\tau_{1;
\alpha}{}^\beta\right]\delta_{nm}.
\label{f9}
\end{equation}

Following \cite{b1} we restrict ourselves to observables in which
the matrix-valued Weyl symbols are proportional to the identity
matrix
\begin{equation}
A_\alpha {}^\beta  (p,q) = A(p,q)\delta _\alpha {}^\beta
\label{f10}
\end{equation}
and call the class of dynamical variables, which
corresponds to such symbols, the class of
 charge-invariant observables.
It is worth noting that the Hamiltonian and current do not belong
to this class. However, in Section \ref{s3} we  present a symbol
that plays the role of the Hamiltonian in our consideration.

The Weyl transformation for charge-invariant observables in the
energy representation has the form
\begin{equation}
A^E {}_{nm;\alpha}{}^\beta= R_\alpha{}^\beta (m,n)\int\limits_{ -
\infty }^{ + \infty } {A(p,q)W_{nm} (p,q)dp\,dq}, \label{f11}
\end{equation}
where $A^E{}_{nm;\alpha}{}^\beta$ is the operator matrix of a
charge-invariant observable in the energy representation. Contrary
to nonlocal theory and the nonrelativistic case, there is a
matrix-valued function $R_\alpha{}^\beta (m,n)$
\begin{equation}
R_\alpha{}^\beta(m,n)
=\sum_{l,k=0}^{\infty}\sum_{\gamma=\pm1}U_{lm;\gamma}{}^{\beta}U^{
- 1}_{nk;\alpha}{}^{\gamma} = \varepsilon
(m,n)\delta_\alpha{}^\beta + \chi (m,n)\tau_{1;\alpha}{}^\beta
\label{f12}
\end{equation}
with even and odd parts expressed through the energy spectrum
(\ref{f6}) as follows:
\begin{equation}
\varepsilon (m,n) = \frac{{E(m) + E(n)}}{{2\sqrt
{E(m)E(n)}}},\label{f13}
\end{equation}
\begin{equation}
 \chi (m,n) = \frac{{E(m) - E(n)}}{{2\sqrt {E(m)E(n)} }}. \label{f14}
\end{equation}
Functions $\varepsilon(m,n)$ and $\chi (m,n)$ play a crucial role
in our consideration. We call them  the $\varepsilon$- and
$\chi$-factors.

The expressions for even $[\hat A]$
and odd $\{ \hat A\}$ parts of the operator of a charge-invariant
observable in terms of its Weyl symbol follow from (\ref{f11})
\begin{equation}
\left[ {A^E } \right]_{nm;\alpha}{}^\beta   = \varepsilon
(m,n)\delta _\alpha{}^\beta  \int\limits_{ - \infty }^{ + \infty }
{A(p,q)W_{nm} (p,q)dp\,dq}\label{f15}
\end{equation}
\begin{equation}
\left\{ {A^E } \right\}_{nm;\alpha}{}^\beta   = \chi(m,n)\tau
_{1;\alpha}{}^\beta  \int\limits_{ - \infty }^{ + \infty }
{A(p,q)W_{nm} (p,q)dp\,dq}.\label{f16}
\end{equation}

One can obtain the inverse formulas connecting Weyl symbols with
the operator, and its even and odd parts
\begin{equation}
A(p,q)\delta _\alpha {}^\beta=(2\pi\hbar)^d\sum_{\gamma =
\pm1}\sum_{m,n=0}^{\infty}{R^{ - 1}{}_\gamma{}^\beta(m,n)A^E
{}_{nm;\alpha} {}^\gamma  W_{mn} (p,q)}, \label{f17}
\end{equation}
\begin{equation}
A(p,q)\delta _\alpha{}^\beta =(2\pi\hbar)^d\sum_{mn}{\varepsilon
^{ - 1} (m,n)\left[ {A^E } \right]_{nm;\alpha}{}^\beta W_{mn}
(p,q)},\label{f18}
\end{equation}
\begin{equation}
A(p,q)\delta _\alpha{}^\beta=(2\pi\hbar)^d\sum_{\gamma =
\pm1}\sum_{m,n=0}^{\infty}{\chi ^{ - 1} (m,n)\tau _{1\gamma}
{}^\beta \left\{ {A^E } \right\}_{nm;\alpha} {}^\gamma  W_{mn}
  (p,q)}.\label{f19}
\end{equation}
Comparing (\ref{f18}) and (\ref{f19}) we conclude that matrix
elements of even and odd parts of the operator of an arbitrary
charge-invariant observable are uniquely related to each other
\begin{equation}
\{A^{E}\} _{nm;\alpha}{}^\beta=\frac{{E(m) - E(n)}}{{E(m) +
E(n)}}\sum_{\gamma=\pm 1} {\tau_{1;\gamma} {}^\beta [A^{E}
]_{nm;\alpha}{}^\gamma}.\label{f20}
\end{equation}

Now we consider the expression for the time derivative of a
charge-invariant observable through its Weyl symbol as was done in
\cite{b1}. To do this, one should use the Heisenberg equation
\begin{equation}
\partial_t\hat{A}=\frac{1}{i\hbar}\left[\hat{A},\hat{H}\right].
\label{f21}
\end{equation}
The commutator can be written in the form of Weyl transform of the
matrix-valued Moyal bracket (see \cite{b1}). Equation (\ref{f21})
in the energy representation can be written as follows
\begin{equation}
\partial_t A^{E}_{nm;\alpha}{}^\beta=\int\limits_{-\infty}^{+\infty}
\left\{A(p,q),G_\alpha{}^\beta\left(m,\{p,q\},n\right)\right\}_M
W_{nm}(p,q)dp\,dq \label{f22}
\end{equation}
where the notation
\begin{equation}
G_\alpha{}^\beta\left(m,\{p,q\},n\right)=\frac{E^{\star
2}(p,q)}{2\sqrt{E(m)E(n)}}\left(\tau_{3;\alpha}{}^\beta+
i\tau_{2;\alpha}{}^\beta\right)
\label{f23}
\end{equation}
\begin{equation}
E^{\star 2}(p,q)=m^2c^4+c^2\Big(p-eA\left(q\right)\Big)^2
\label{f24}
\end{equation}
is introduced.

The Moyal bracket is defined in such a way to provide coincidence
with the Poisson bracket at $\hbar\rightarrow 0$, namely,
\begin{eqnarray}
\left\{A(p,q),B(p,q)\right\}_M&=\frac{1}{i\hbar}\left(A(p,q)\star
B(p,q)-B(p,q)\star A(p,q)\right)\nonumber \\
&=\frac{2}{\hbar}A(p,q)\sin\left\{\frac{\hbar}{2}\left(
\overleftarrow{\partial_q}\overrightarrow{\partial_p}-
\overleftarrow{\partial_p}\overrightarrow{\partial_q}\right)
\right\}B(p,q)\label{f25}
\end{eqnarray}
where the Moyal star-product is determined as usually
\begin{equation}
\star\equiv\exp\left\{\frac{i\hbar}{2}\left(
\overleftarrow{\partial_q}\overrightarrow{\partial_p}-
\overleftarrow{\partial_p}\overrightarrow{\partial_q}\right)\right\}.
\label{f26}
\end{equation}

Contrary to the free particle case~\cite{b1}, expression
(\ref{f22}) does not have a classical form even if $A(p,q)$ is a
linear function of $p$ and $q$. The semiclassical approximation in
static magnetic field for the standard theory has a more
complicated structure than that used in nonlocal theory. In
Section \ref{s4} we consider another approach to the Heisenberg
picture of motion and show that these peculiarities are related to
the form of the Wigner function only and do not change the
equation of motion.

\section{Wigner function and quantum Liouville equation
for charge-invariant observables}
\label{s3}

In order to find the explicit form of the Wigner function for
charge-invariant observables, we start with the expression for
means. Let $C_{n;\alpha}$ be the wave function in the energy
representation (the decomposition coefficient of the wave function
presented in the form of series in eigenstates of the Hamiltonian
(\ref{f3})). Thus the mean of the charge-invariant observable
$\hat{A}$ is written in the explicit form
\begin{equation}
\overline{A}=\sum_{\alpha,\beta=\pm1}\sum_{m,n=0}^{\infty}C_{m;\beta}
^{*}C_{n;}{}^\alpha
 A^{E}_{nm;\alpha}{}^\beta. \label{f27}
\end{equation}
Substituting (\ref{f11}) in (\ref{f27}), one can obtain for the
mean
\begin{equation}
\overline{A}=\int\limits_{-\infty}^{+\infty}A(p,q)W(p,q)dp\,dq,
\label{f28}
\end{equation}
where the Wigner function for charge-invariant observable can be
presented as the sum of four terms
\begin{equation}
W(p,q)=\sum_{\alpha=\pm 1
}\left(W_{[\alpha]}(p,q)+W_{\{\alpha\}}(p,q)\right). \label{f29}
\end{equation}
This approach has been applied to construct the
covariant Wigner function in \cite{b5}. We call $W_{[\pm]}(p,q)$
 the
 even part of the Wigner function. It can be written as follows
 \begin{equation}
W_{[\pm]}(p,q) =\sum_{n,m=0}^{\infty}{\varepsilon(m,n)} W_{nm}
(p,q)C_{m;}^*{}_{\pm}C_{n;}{}^\pm. \label{f30}
\end{equation}
In a similar way, we call $W_{\{\pm\}}(p,q)$  the odd part of the
Wigner function defined as
\begin{equation}
W_{\{\pm\}}(p,q) =\sum_{n,m=0}^{\infty}{\chi(m,n)} W_{nm}
(p,q)C_{m;}^*{}_{\pm}C_{n;}{}^{\mp}. \label{f31}
\end{equation}
Note that here use other symbols for the even and odd parts of the
Wigner function than those employed in \cite{b1}.

The Wigner function for charge-invariant observables can be
defined in an alternative way. To do this, we consider the wave
function in the nonlocal theory representation
\begin{equation}
\psi _\pm(p)=\sum_{n=0}^{\infty}C_{n;\pm}\varphi_n(p). \label{f32}
\end{equation}
Now one can write the components of the Wigner function as follows
\begin{eqnarray}
W_{[\pm]}(p,q)\!=\!\frac{1}{{(2\pi \hbar )^d
}}\!\int\limits_{-\infty}^{+\infty}\! {\varepsilon\! \left(\!p\!
+\! \frac{P}{2}\!,\!p_1\!;\!p\! -\! \frac{P}{2}\!,\!p_2\!
\right)\!\psi _\pm ^ *\! (p_1 )\!\psi ^\pm\! (p_2 )\!e^{-
\frac{i}{\hbar}Pq}\! d\!P\,d\!p_1\,d\!p_2 }\!,\nonumber\\
\label{f33}\\ W_{\{\pm\}}(p,q)\!=\!\frac{1}{{(2\pi \hbar )^d
}}\!\int\limits_{-\infty}^{+\infty}\! {\chi\! \left(\!p\! +\!
\frac{P}{2}\!,\!p_1\!;\!p\! -\! \frac{P}{2}\!,\!p_2\!
\right)\!\psi _\pm ^ *\! (p_1 )\!\psi ^\mp\! (p_2 )\!e^{-
\frac{i}{\hbar}Pq}\! d\!P\,d\!p_1\,d\!p_2 }\!,\nonumber\\
\label{f34}
\end{eqnarray}
where the following generalized functions are introduced:
\begin{equation}
\varepsilon (p^\prime,p_1 ;p^{\prime\prime},p_2 ) =
\sum_{n,m=0}^{\infty} {\varepsilon (m,n)\varphi _m^ *
(p^\prime)\varphi _m (p_1 )\varphi _n (p^{\prime\prime})\varphi
_n^
*  (p_2 )} \label{f35}
\end{equation}
\begin{equation}
\chi (p^{\prime},p_1 ;p^{\prime\prime},p_2 ) =
\sum_{n,m=0}^{\infty} {\chi (m,n)\varphi _m^
* (p^\prime)\varphi _m (p_1 )\varphi _n (p^{\prime\prime})\varphi _n^*
(p_2
)}.\label{f36}
\end{equation}

Following \cite{b1} we obtain the evolution equation for each
component separately. To do this, we differentiate expressions
(\ref{f30}) and (\ref{f31}) with respect to time and substitute
the time derivatives from the Klein--Gordon equation in the energy
representation
\begin{equation}
i\hbar\partial_t
C_{n;\alpha}=E(n)\sum_{\beta=\pm1}\tau_{3;}{}^\beta{}_\alpha
C_{n;\beta}. \label{f37}
\end{equation}
In view of the star-eigenvalue equations \cite{b14}
\begin{eqnarray}
E(p,q)\star W_{nm}(p,q)=E(n) W_{nm}(p,q)\nonumber \\
W_{nm}(p,q)\star E(p,q)=E(m) W_{nm}(p,q) \label{f38}
\end{eqnarray}
we obtain the quantum Liouville equation in the following form:
\begin{eqnarray}
\partial_tW_{[\pm]}(p,q,t)
=\pm\left\{E(p,q),W_{[\pm]}(p,q,t)\right\}_M \label{f39}\\
\partial_tW_{\{\pm\}}(p,q,t)=\mp\left[E(p,q),W_{\{\pm\}}(p,q,t)\right]_M.
\label{f40}
\end{eqnarray}
The Moyal bracket in (\ref{f39}) is defined by expressions
(\ref{f25}). Furthermore, we have used the anti-Moyal bracket in
(\ref{f40}), which is defined as follows
\begin{eqnarray}
\left[A(p,q),B(p,q)\right]_M&=\frac{1}{i\hbar}\Big(A(p,q)\star
B(p,q)+B(p,q)\star A(p,q)\Big)\nonumber \\
&=\frac{2}{i\hbar}A(p,q)\cos\left\{\frac{\hbar}{2}\left(
\overleftarrow{\partial_q}\overrightarrow{\partial_p}-
\overleftarrow{\partial_p}\overrightarrow
{\partial_q}\right)\right\}B(p,q).\label{f41}
\end{eqnarray}

In these formulas,$E(p,q)$ plays the role of the Hamiltonian; it
is determined as `the star square root' of expression (\ref{f24})
\begin{equation}
E(p,q) = \sqrt[\star]{{m^2 c^4  + c^2 [p - eA(q)]^2}} .
\label{f42}
\end{equation}
This is fairly unexpected, since in
 nonrelativistic
quantum mechanics a classical variable is mapped onto the
corresponding Weyl symbol (at least, in the Schr\"odinger picture
of motion or at the initial time moment in the Heisenberg
picture). The classical Hamilton function differs from (\ref{f42})
since it employs the square root instead of the star square root.
It gives different pictures of the classical and quantum
evolutions.

In the Appendix, we consider an approximate expression for the
relativistic rotator (simplified model of the particle motion
 in a
 homogeneous magnetic field). From this example,
 one can see that
 the star square root results in the appearance
 of independent power series expansion in the cyclotron frequency
together with
 the expansion in position and momentum (\ref{af16}). The
cyclotron frequency is inversely proportional to the square of the
characteristic oscillator length, which corresponds, in many
cases, to the wave packet width. Therefore, one can say, that the
star square root introduced results in specific effects which
appear in strong magnetic field
 and for very strong localization (of the order
of the Compton wavelength).

It is worth noting that such behaviour of the relativistic system
does not relate to the nontrivial charge structure of position and
momentum operators because  equation (\ref{f39}) with Hamiltonian
(\ref{f42}) are valid for nonlocal theory as well.

Consider the odd part of the Wigner function (\ref{f31}),
(\ref{f34}). It was mentioned in \cite{b5}, that it is not
important for macroscopic values. Indeed, the existence of states,
where this part of the Wigner function is not equal to zero
contradicts to charge superselection rule~\cite{b15}. Such a state
is equivalent to the superposition of states with different charge
 signs. According to the contemporary concept, if such state
appears as a result of the physical process, the superposition is
reduced to the mixed state, which results in the creation of
particle pairs~\cite{b16}. In our representation, this means that
the odd part of the Wigner function becomes zero. However, it
appears when an external electric or nonstatic field is applied.
Therefore, we take into account its existence here.

The solutions of the system  of equations (\ref{f39}) and
(\ref{f40}) are not
 independent. There exists a specific constraint on them.
 To find it, one
 should take the Fourier transform and make the standard
change of variables in each component of (\ref{f33}), (\ref{f34}).
Similar to the free particle case \cite{b1}, we obtain four
expressions where on the left-hand side there are products of two
wave functions of different arguments, and on the right-hand side
we obtain integral expressions dependent on the Wigner function
components. Splitting them into pairs and equating the resulting
expressions, due to the equality of their left-hand sides, we
obtain the constraint we are looking for:
\begin{eqnarray}
\int\limits_{ - \infty }^{ + \infty } {\varepsilon ^{ - 1} (p_1
,p^\prime;p_2 ,p^{\prime\prime})W_{[+]} \left(\frac{1}{2}(p_1 +
p_2 ),q_1 \right)e^{\frac{i}{\hbar }(p_1 - p_2 )q} dp_1\,dp_2
}\,dq\nonumber
\\ \times\int\limits_{ - \infty }^{ + \infty } {\varepsilon ^{ -
1} (p_1 ,p^\prime;p_2 ,p^{\prime\prime})W_{[-]}
\left(\frac{1}{2}(p_1 + p_2 ),q_2 \right)e^{\frac{i}{\hbar }(p_1 -
p_2 )q} dp_1\,dp_2\,dq}\nonumber \\ = \int\limits_{ - \infty }^{ +
\infty } {\chi ^{ - 1} (p_1 ,p^\prime;p_2
,p^{\prime\prime})W_{\{+\}} \left(\frac{1}{2}(p_1 + p_2 ),q_1
\right)e^{\frac{i}{\hbar }(p_1 - p_2 )q}
dp_1\,dp_2}\,dq\nonumber\\ \times\int\limits_{ - \infty }^{ +
\infty } {\chi ^{ - 1} (p_1 ,p^\prime;p_2
,p^{\prime\prime})W_{\{-\}} \left(\frac{1}{2}(p_1  + p_2 ),q_2
\right)e^{\frac{i}{\hbar }(p_1  - p_2 )q}
dp_1\,dp_2\,dq},\nonumber\\ \label{f43}
\end{eqnarray}
where the following generalized functions are introduced:
\begin{equation}
\varepsilon ^{ - 1} (p_1 ,p^\prime;p_2 ,p^{\prime\prime}) =
\sum_{n,m=0}^{\infty} {\varepsilon ^{ - 1} (m,n)\varphi _m^ *
(p^\prime)\varphi _m (p_1 )\varphi _n (p^{\prime\prime})\varphi
_n^* (p_2 )} \label{f44}
\end{equation}
\begin{equation}
\chi ^{ - 1} (p_1 ,p^\prime;p_2 ,p^{\prime\prime}) =
\sum_{n,m=0}^{\infty} {\chi ^{ - 1} (m,n)\varphi _m^ *
(p^\prime)\varphi _m (p_1 )\varphi _n (p^{\prime\prime})\varphi
_n^*  (p_2 )}. \label{f45}
\end{equation}

In addition to this constraint, there exist expressions for
complex conjugate values of the Wigner function, which can be
considered as specific constraints as well. To obtain them, one
should take the complex conjugate of (\ref{f30}), (\ref{f31}) or
(\ref{f33}), (\ref{f34}). After some algebra, we obtain the
following expressions:
\begin{equation}
W^*_{[\pm]}(p,q) = W_{[\pm]}(p,q)  \label{f46}
\end{equation}
\begin{equation}
W^*_{\{\pm\}}(p,q) = W_{\{\mp\}}(p,q).\label{f47}
\end{equation}
The even components of the Wigner function are real functions and
the odd
 components are complex conjugate to each other; therefore,
their sum
 is the real function, as well.

\section{Heisenberg picture of motion}
\label{s4}

It is easy to map the Schr\"odinger picture onto the Heisenberg
picture of motion and vice versa in the nonrelativistic WWM
formalism. In our case, this map has some peculiarities due to the
existence of different equations for the odd and even parts of the
Wigner function.

Let us write the expression for the mean of a
charge-invariant observable (\ref{f28}) as a function of time,
taking into account the structure of the Wigner function
(\ref{f29})
\begin{equation}
\overline{A}(t)=\sum_{\alpha=\pm 1
}\int\limits_{-\infty}^{+\infty}A(p,q)
\left(W_{[\alpha]}(p,q,t)+W_{\{\alpha\}}(p,q,t)\right)dp\,dq.
\label{f48}
\end{equation}
Now we extend the definition of the Weyl symbol for a
charge-invariant observable. To do this, we introduce four new
symbols $A_{[\pm]}(p,q)$, $A_{\{\pm\}}(p,q)$, which coincide with
$A(p,q)$ in the Schr\"odinger picture of motion. Expression
(\ref{f48}) can be written as the sum of four components:
\begin{equation}
\overline{A}(t)=\sum_{\alpha=\pm 1
}\left(\overline{A}_{[\alpha]}(t)+
\overline{A}_{\{\alpha\}}(t)\right),\label{f49}
\end{equation}
\begin{equation}
\overline{A}_{[\alpha]}(t)=\int\limits_{-\infty}^{+\infty}
A_{[\alpha]}(p,q)W_{[\alpha]}(p,q,t)\,dp\,dq, \label{f50}
\end{equation}
\begin{equation}
\overline{A}_{\{\alpha\}}(t)=
\int\limits_{-\infty}^{+\infty}A_{\{\alpha\}}(p,q)W_{\{\alpha\}}
(p,q,t)\,dp\,dq
. \label{f51}
\end{equation}
We consider the map onto the Heisenberg picture of
motion for each component separately.

The formal solution of  equations (\ref{f39}) and (\ref{f40}) can
be written as follows:
\begin{eqnarray}
W_{[\pm]}(p,q,t)=\exp_{\star}\left(\mp\frac{i}{\hbar}H(p,q)t\right)\star
W_{[\pm]}(p,q)\star\exp_{\star}\left(\pm\frac{i}{\hbar}H(p,q)t\right),
\label{f52}\\
W_{\{\pm\}}(p,q,t)=\exp_{\star}\left(\pm\frac{i}{\hbar}H(p,q)t\right)\star
W_{\{\pm\}}(p,q)\star\exp_{\star}\left(\pm\frac{i}{\hbar}H(p,q)t\right),
\label{f53}
\end{eqnarray}
where the exponent is determined by means of the star-product
\cite{b13}. Substituting this solution into (\ref{f50}) and
(\ref{f51}) and writing this expression in the standard operator
form, one obtains
\begin{eqnarray}
\overline{A}_{[\pm]}(t)=\Tr\left\{\hat{A}_{[\pm]}
\exp\left(\mp\frac{i}{\hbar}\hat{H}t\right)
\hat{\rho}_{[\pm]}\exp\left(\pm\frac{i}{\hbar}\hat{H}t\right)
\right\},\label{f54}\\
\overline{A}_{\{\pm\}}(t)=\Tr\left\{\hat{A}_{\{\pm\}}
\exp\left(\pm\frac{i}{\hbar}\hat{H}t\right)
\hat{\rho}_{\{\pm\}}\exp\left(\pm\frac{i}{\hbar}\hat{H}t\right)\right\}.
\label{f55}
\end{eqnarray}

Making the cyclic permutation under the trace and returning into
the WWM representation, one can write  expressions (\ref{f50}) and
(\ref{f51}) as follows
\begin{equation}
\overline{A}_{[\alpha]}(t)=
\int\limits_{-\infty}^{+\infty}A_{[\alpha]}(p,q,t)W_{[\alpha]}(p,q)
\,dp\,dq, \label{f56}
\end{equation}
\begin{equation}
\overline{A}_{\{\alpha\}}(t)=\int\limits_{-\infty}^{+\infty}
A_{\{\alpha\}}(p,q,t)W_{\{\alpha\}}(p,q)\,dp\,dq,
 \label{f57}
\end{equation}
with the time-dependent symbol of the
charge-invariant observable
\begin{eqnarray}
A_{[\pm]}(p,q,t)=\exp_{\star}\left(\pm\frac{i}{\hbar}H(p,q)t\right)\star
A_{[\pm]}(p,q)\star\exp_{\star}\left(\mp\frac{i}{\hbar}H(p,q)t\right),
\label{f58}\\
A_{\{\pm\}}(p,q,t)=\exp_{\star}\left(\pm\frac{i}{\hbar}
H(p,q)t\right)\star
 A_{\{\pm\}}(p,q)\star\exp_{\star}
\left(\pm\frac{i}{\hbar}H(p,q)t\right).
\label{f59}
\end{eqnarray}
It is easy to see that this symbol satisfies the following
equations:
\begin{eqnarray}
\partial_tA_{[\pm]}(p,q,t)=\pm\left\{A_{[\pm]}(p,q,t),E(p,q)
\right\}_M, \label{f60}\\
\partial_tA_{\{\pm\}}(p,q,t)=\mp\left[A_{\{\pm\}}(p,q,t),E(p,q)
\right]_M.
\label{f61}
\end{eqnarray}

Therefore, we conclude that the time evolution for
charge-invariant observables is identical in  both the standard
and nonlocal theories under conditions where the creation of
particle pairs does
 not take place. The difference is in the $\varepsilon$-factor
in the
 definition of the Wigner function that affects the
class of possible functions that describes the real physical
state.

\section{Statistical properties of the Wigner function
for charge-invariant observables}
\label{s5}

The Wigner function is expressed in different ways in nonlocal
theory  and in the standard theory. Therefore, one can expect
 some peculiarities for the case of standard theory.
The evolution equations are the
 same in both theories for
the systems with stable vacuum. However, it is well known
\cite{b16a} that a difference of the WWM formalism in
nonrelativistic quantum mechanics (and in nonlocal theory as well)
and classical mechanics is the particular constraint on the Wigner
function. In this Section, we show that in our case this
constraint has some peculiarities and, generally speaking, differs
from the one used in  nonlocal (and nonrelativistic) theory. This
results in some peculiarities in the mean values of
charge-invariant observables. Let us start with usual properties
of the distribution function.

\begin{property}[normalization] \label{p1}
The even part of the Wigner function (\ref{f30}), (\ref{f33}) is
normalized in the whole phase space, and integral of the odd part
(\ref{f31}), (\ref{f34}) is equal to zero.
\end{property}

The distribution function of the momentum (position) for  a charge
definite state (we do not take into account a superposition of
states with different charge signs) is  the integral over the
position $q$ (momentum $p$) of the even part of the Wigner
function (\ref{f30}), (\ref{f33}).

\begin{property}\label{p2}
The distribution functions of the momentum and position can be
written as follows
\begin{equation}
\rho_{\pm}(p) = \psi _\pm ^ *  (p)\varepsilon \left(\hat
{\overleftarrow{n}},\hat {\overrightarrow{n}}\right)\psi ^\pm
(p),\label{f62}
\end{equation}
\begin{equation}
\rho_{\pm}(q) = \psi _\pm ^ *  (q)\varepsilon \left(\hat
{\overleftarrow{n}},\hat {\overrightarrow{n}}\right)\psi ^\pm
(q)\label{f63}
\end{equation}
where $\psi _\pm(q)$, $\psi _\pm(p)$ are the wave functions in the
nonlocal theory representation. Symbols $\hat {\overleftarrow{n}}$
and $\hat {\overrightarrow{n}}$ correspond to operators acting on
the function from the left and right sides, and the number $n$ in
(\ref{f6}), (\ref{f13}) is their eigenvalue .
\end{property}

The distribution functions (\ref{f62}), (\ref{f63}) differ from
those used in nonlocal theory. Furthermore, they can take negative
values. This specific property of the spin-0 particle described by
the Klein--Gordon equation creates difficulties in the probability
interpretation. We do not discuss this feature here, assuming this
anomaly as a disease of the model (for details and a review of the
problem see, for example, \cite{b2, b2a,b3,b7} and references
therein). However, it is worth noting, that for spin-1/2 particles
the problem is absent though the nontrivial charge structure of
the position and momentum operators exists here as well
\cite{b8b}.

The peculiarity of distributions (\ref{f62}), (\ref{f63}) results
in nontrivial mean values of charge-invariant observables. We
consider this on the example of moments of the position and
momentum.

\begin{property}\label{p3}
The $n$th moment of the position and momentum can be written as
follows
\begin{equation}
\overline{q^n}= \sum_{l,m=0}^{\infty} {\left(q^n\right)_{lm}
C_{m;\pm} ^
* \varepsilon(m,l) C_{l;}{}^\pm},\label{f64}
\end{equation}
\begin{equation}
\overline{p^n}= \sum_{l,m=0}^{\infty} {\left(p^n\right)_{lm}
C_{m;\pm} ^
* \varepsilon(m,l) C_{l;}{}^\pm},\label{f65}
\end{equation}
where $\left(q^n\right)_{lm}$ and $\left(p^n\right)_{lm}$ are the
matrix elements for the $n$th power of position and momentum
operators in nonlocal theory.
\end{property}

The first moments of the position and momentum coincide with those
used in  nonlocal theory for the free particle case~\cite{b1}. It
is not true, generally speaking, for the case of arbitrary
magnetic field. Due to this, there exist peculiarities in the
behaviour of the mean values of positions and momenta, i.e.
 the peculiarities
for the mean trajectories. This problem will be considered in
detail elsewhere.

Now, similar to \cite{b16a}, we formulate the property that can be
considered as a constraint on the Wigner function for
charge-invariant observables. This also enables us to select from
the set of functions on the phase space those that can be
considered as the Wigner functions for charge-invariant
observables.

\enlargethispage{\baselineskip}
\begin{property}[criterium of pure state] \label{p4}
For the functions $W_{[\pm]}(p,q)$ and $W_{\{\pm\}}(p,q)$ to be
even and odd components of the Wigner function for
charge-invariant observables, it is necessary and sufficient that
equalities (\ref{f43}), (\ref{f46}), (\ref{f47}) hold true and the
following conditions are satisfied:
\begin{eqnarray}
\frac{{\partial ^2 }}{{\partial p_1 \partial p_2 }}\!\ln\!
\int\limits_{ - \infty }^{ + \infty }\!{\varepsilon ^{- 1}
(p^{\prime},p_1 ;p^{\prime\prime},p_2
)W_{[\pm]}\!\left(\frac{1}{2}\left(p^{\prime}\! +\!
p^{\prime\prime}\right)\!,\!q\right)\!e^{\frac{i}{\hbar }(p^\prime
- p^{\prime\prime})q} d\!q\,d\!p^\prime d\!p^{\prime\prime}\! =\!
0}\nonumber\\ \label{f66}\\  \frac{{\partial ^2 }}{{\partial p_1
\partial p_2 }}\!\ln\! \int\limits_{ - \infty }^{ + \infty }\! {\chi^{-
1} (p^{\prime},p_1 ;p^{\prime\prime},p_2
)W_{\{\pm\}}\!\left(\frac{1}{2}\left(p^{\prime}\! +\!
p^{\prime\prime}\right)\!,\!q\right)\!e^{\frac{i}{\hbar }(p^\prime
- p^{\prime\prime})q} d\!q\,d\!p^\prime d\!p^{\prime\prime}\! =\!
0}\nonumber
\\ \label{f67}
\end{eqnarray}
where $\varepsilon ^{- 1} (p^{\prime},p_1 ;p^{\prime\prime},p_2 )$
and $\chi^{- 1} (p^{\prime},p_1 ;p^{\prime\prime},p_2 )$ are the
generalized functions determined by (\ref{f44}), (\ref{f45}).
\end{property}

The proof of this criterium is similar to that used for the free
particle case~\cite{b1}. One should use here the expression for
the Wigner function in view of the wave function in the nonlocal
theory representation (\ref{f33}), (\ref{f34}).

Property \ref{p4} with equations (\ref{f39}), (\ref{f40}) in the
Schr\"odinger picture or (\ref{f60}), (\ref{f61}) in the
Heisenberg picture of motion are sufficient to formulate the
quantum problem in the representation used. Furthermore,
Property~\ref{p4} is a characteristics that distinguishes the
standard theory from the nonlocal theory under conditions where
the creation of particle pairs is not possible. However, there
exist examples, where the Wigner function is the same in both the
standard and nonlocal theories.

\begin{property} \label{p5}
The Wigner function for charge-invariant observables which
describes a stationary state coincides with that used in nonlocal
theory.
\end{property}

It can be an eigenstate of the Hamiltonian or mixture of such
states. This Property is the consequence of the definition of the
Wigner function for charge-invariant observables (\ref{f30}),
 (\ref{f31}) and the obvious fact
that $\varepsilon(n,n)=1$. Hence, the peculiarities related to the
Properties \ref{p2}--\ref{p4} do not appear in  stationary
processes (for example, in equilibrium statistical physics).

\section{Physical meaning, simple examples and possibility of
 experimental observation}
\label{s6}

The $\varepsilon$-factor (\ref{f13}) plays a crucial role in our
consideration and the problem of its physical meaning is
very important. To clarify it, let us consider some properties of the
$\varepsilon$-factor as a function of two arguments:
\begin{enumerate}
\item\label{ii1}
Symmetry, $\varepsilon(n,m)=\varepsilon(m,n)$;
\item\label{ii2}
Value of diagonal elements, $\varepsilon(n,n)=1$;
\item\label{ii3}
Value of nondiagonal elements, $\varepsilon(n,m)>1$, if $n \neq
m$.
\end{enumerate}
This function is plotted in figure \ref{fig1} for the case of a
relativistic rotator (see Appendix) and for a free particle. The
$\varepsilon$-factor is a slowly increasing function on the both
sides of the diagonal.

\begin{figure}
\begin{center}
\mbox{\includegraphics[width=0.46\textwidth, clip=]{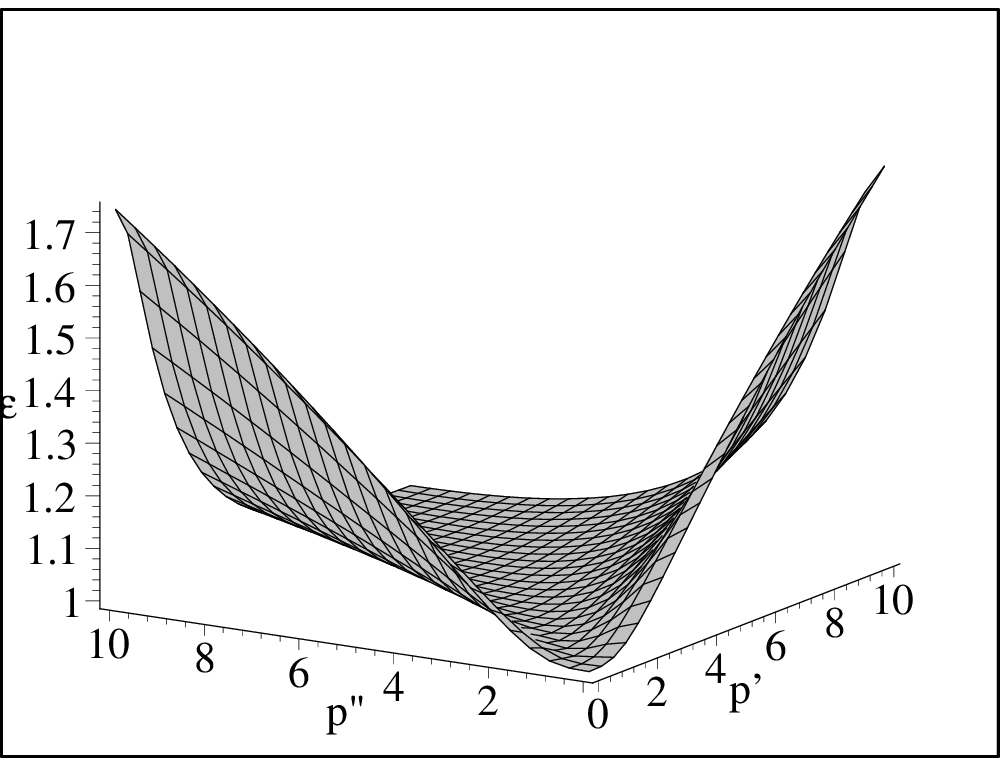}
{a}\includegraphics[width=0.46\textwidth, clip=]{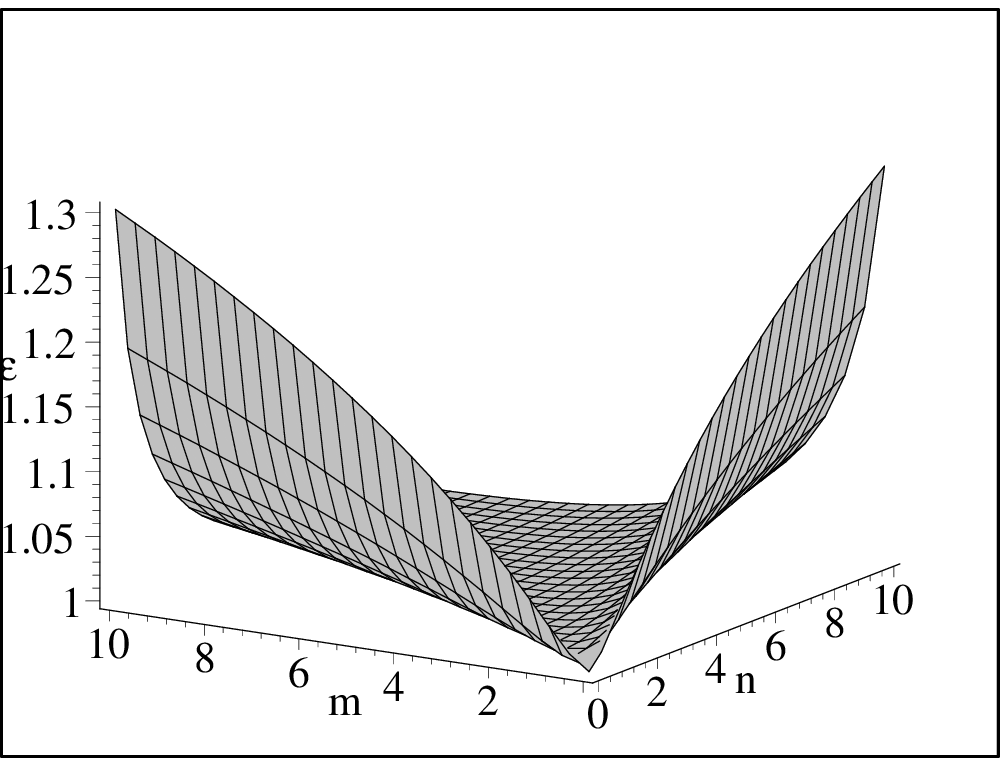}{b} }
\end{center}
\caption{\label{fig1} The $\varepsilon$-factor for a free particle
where momenta are given in $mc$ units~(a) and for a relativistic
rotator where the
 ratio of the Compton wavelength
to the oscillator length is
 $\lambda=10$~(b).}
\end{figure}

Taking into account (\ref{ii2}),  expression (\ref{f30}) for the
even part of the Wigner function can be rewritten as follows:
\begin{equation}
W_{[\pm]}(p,q) =\pm\sum\limits_{n=0}^{\infty} W_{nn}
(p,q)\left|C_{n;\pm}\right|^2+\sum\limits_{n\neq
m}{\varepsilon(m,n)} W_{nm} (p,q)C_{m;}^*{}_{\pm}C_{n;}{}^\pm.
\label{f68}
\end{equation}
Therefore, the $\varepsilon$-factor affects the value of the
interference terms only. Taking into account (\ref{ii3}), one can
say, that it results in an effective increase in  coherence
between eigenstates of the Hamiltonian. Hence, information on the
relative phase between $C_{n;\pm}$ becomes more evident; moreover,
the value of this phase does not change.

It is worth noting, that this information can be lost. This is
possible in the case where the particle interacts with an
environment (or has interacted with it in the past) in such a way
that they are in entangled state
\begin{equation}
\left|\Psi\right\rangle=\sum_{n=0}^{\infty}C_n\left|n\right\rangle
\otimes\left|a(n)\right\rangle, \label{f69}
\end{equation}
where $\left|n\right\rangle$ are the particle eigenstates and
$\left|a(n)\right\rangle$ are  macroscopically distinguishable
states (generally not orthogonal) of an environment, which
includes many degrees of freedom. In this and other equations, we
do not write explicitly the symbols $\pm$.

The problem of decoherence is one of the most important in quantum
information and quantum technologies (see, for example,
\cite{b16a1} and references therein). Also, this problem has a
fundamental meaning because it plays a crucial role in
understanding the quantum measurement processes. We consider
briefly this problem to clarify the possibility of using  the
nontrivial charge structure of the position (momentum) operator
for suppressing the decoherence.

The density operator of the particle, considered as an open
system, is the operator obtained after averaging of the pure state
$\left|\Psi\right\rangle\left\langle\Psi\right|$
over the degrees of freedom of the environment
\begin{equation}
\hat{\rho}=\sum_{n,m=0}^{\infty}\left|n\right\rangle C_{m}^*C_{n}
a(m,n)\left\langle m\right|, \label{f70}
\end{equation}
where
\begin{equation}
a(m,n)=\left\langle a(m)|a(n)\right\rangle. \label{f71}
\end{equation}

The Wigner function of this state in the nonlocal
(nonrelativistic) theory can be written in the following form:
\begin{equation}
W(p,q) =\sum_{n,m=0}^{\infty}{a(m,n)} W_{nm} (p,q)C_{m}^*C_{n}.
\label{f72}
\end{equation}
This expression is very similar to (\ref{f30}) for
the Wigner function for charge-invariant observables in the
standard theory; $a(m,n)$ plays a role of the
$\varepsilon$-factor here. From (\ref{f71}) the following
properties of $a(m,n)$ follow:
\begin{enumerate}
\item\label{iii1}
Hermiticity, $a^*(n,m)=a(m,n)$;
\item\label{iii2}
Value of diagonal elements, $a(n,n)=1$;
\item\label{iii3}
Absolute value of nondiagonal elements, $\left|a(n,m)\right|<1$,
if $n \neq m$.
\end{enumerate}

Properties (\ref{iii1}) and (\ref{iii2}) are identical to those of
the $\varepsilon$-factor. However, as follows from (\ref{iii3}),
the contribution of interference terms in the open system
decreases. This is essence of the decoherence process. The
$\varepsilon$-factor leads to the opposite result.

Consider the Wigner function for charge-invariant observables that
represents the state of open system (\ref{f70})
\begin{equation}
W_{[\pm]}(p,q) =\sum_{n,m=0}^{\infty}{a(m,n)\varepsilon(m,n)}
W_{nm} (p,q)C_{m;}^*{}_{\pm}C_{n;}{}^\pm. \label{f73}
\end{equation}
Due to the $\varepsilon$-factor, a contribution of interference
terms increases here. Therefore, the nontrivial charge structure
of the position (momentum) operators results in an effective
increase in coherence between eigenstates of the Hamiltonian.

In figure \ref{fig2} we present a plot of the Wigner function for
the state of a relativistic rotator (see Appendix), which is the
superposition of
 two eigenstates of the Hamiltonian,
$\left|0\right\rangle$ and
$\left|2\right\rangle$, for three different cases. Figure
\ref{fig2}~a shows the Wigner function for the mixed
state.
 Figure \ref{fig2}~b corresponds
to the
 superposition in nonlocal theory (without decoherence).
The contribution of interference terms appears in this example.
Figure \ref{fig2}~c shows the Wigner function for charge-invariant
observables in the standard theory. Interference terms play a more
crucial role here.

\begin{figure}
\begin{center}
\mbox{\includegraphics[width=0.29\textwidth, clip=]{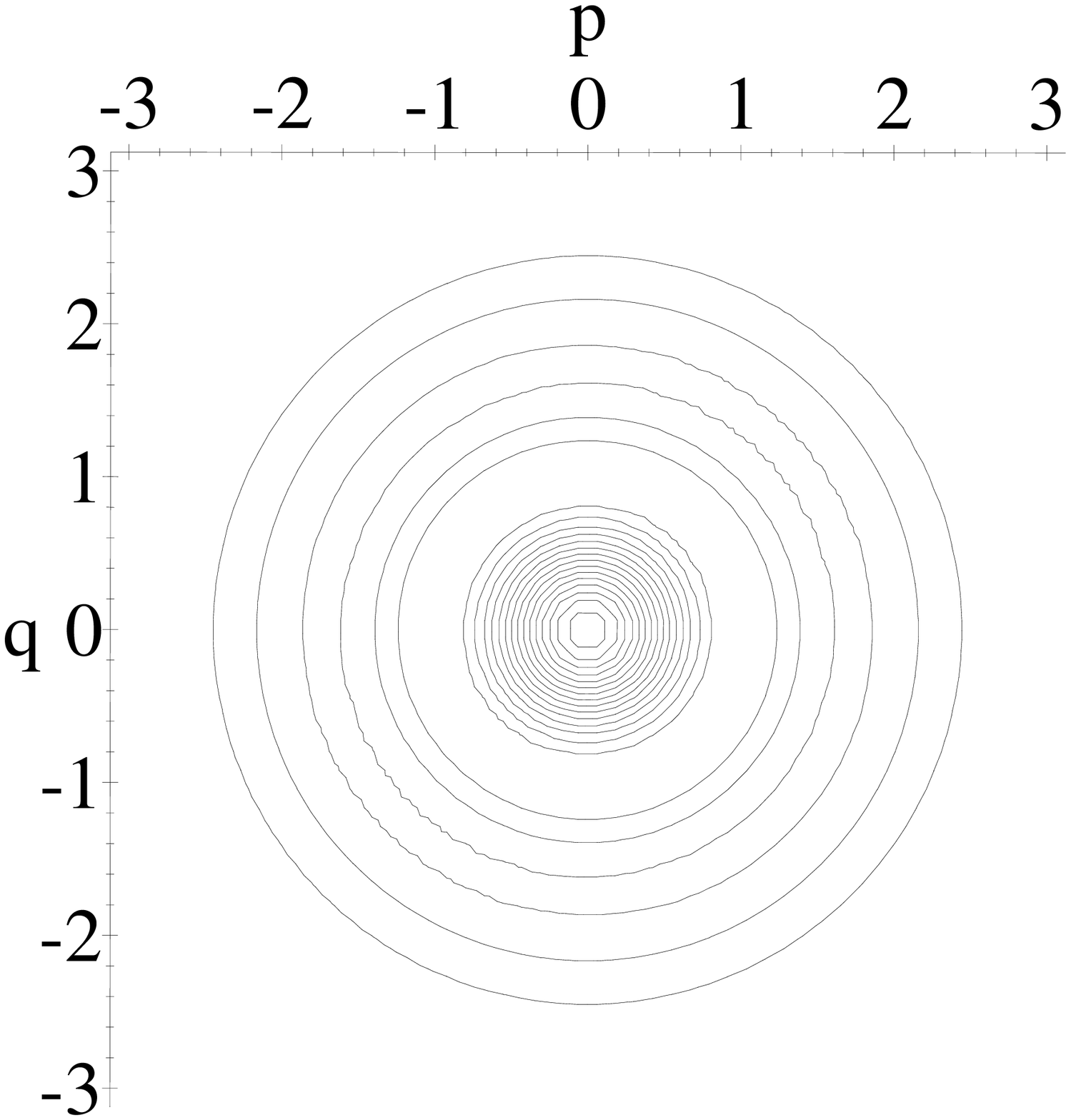}{a}
\includegraphics[width=0.29\textwidth, clip=]{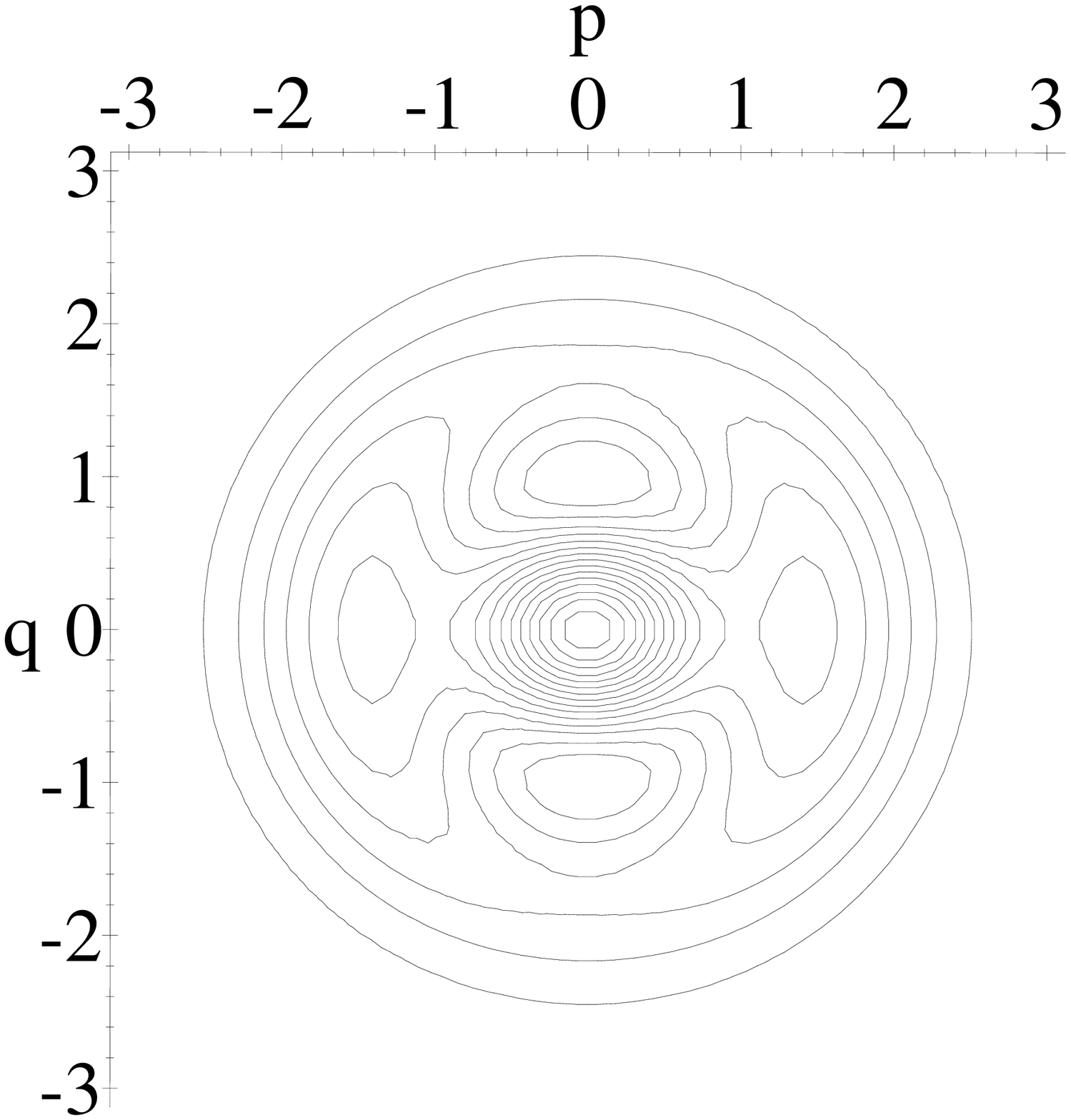}{b}
\includegraphics[width=0.29\textwidth, clip=]{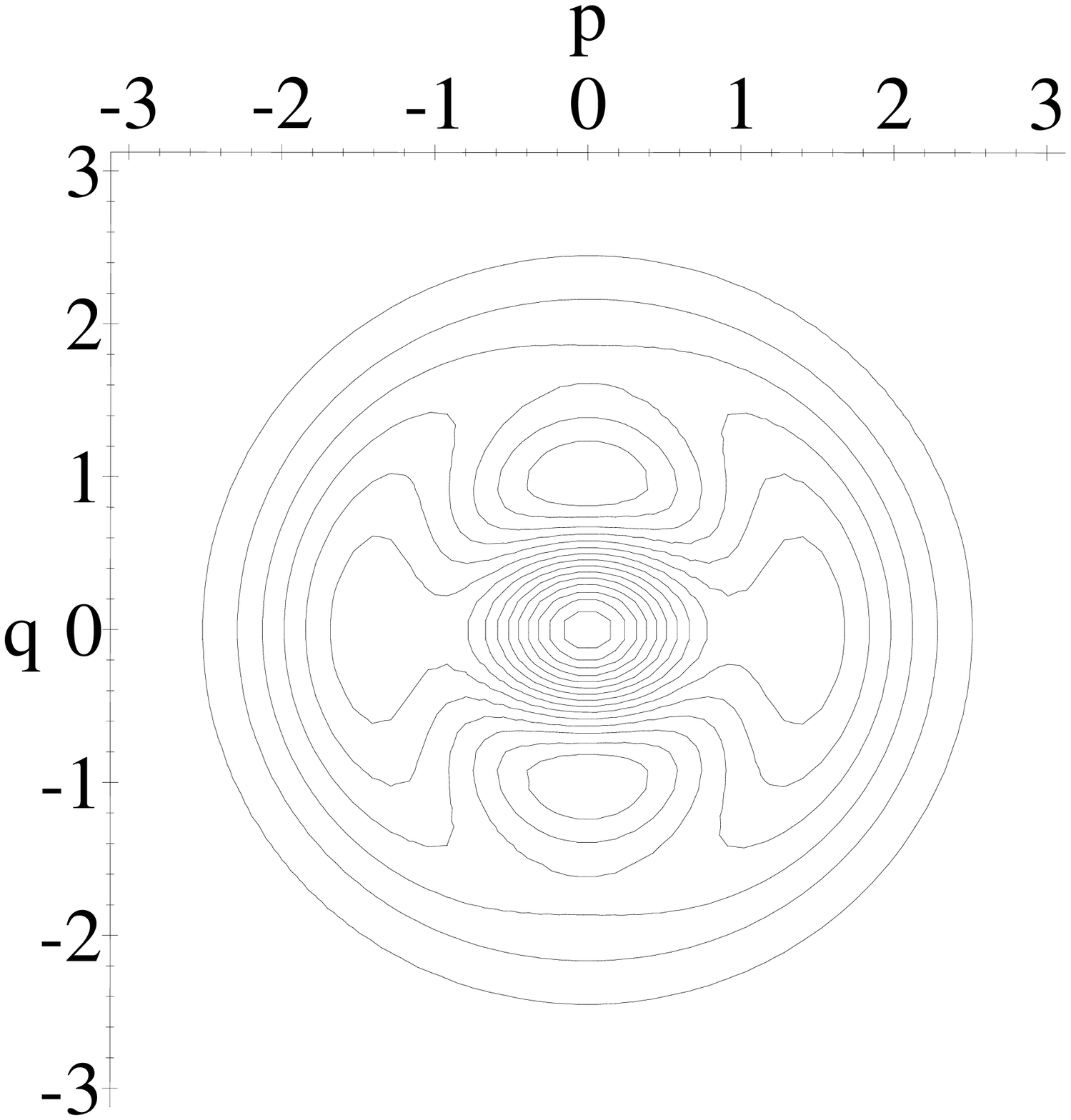}{c}}
\end{center}
\caption{\label{fig2} Contours of the Wigner function of a
relativistic rotator for the mixed state
$\hat{\rho}=1/2\Big(\left|0\right\rangle\left\langle0\right| +
\left|2\rangle\langle 2\right|\Big)$~(a) and for the
 superposition
$\left|\psi\right\rangle=2^{-1/2}\Big(\left|0\right\rangle+
\left|2\right\rangle\Big)$ ~(b) in nonlocal theory and in the
standard theory with nontrivial charge structure of the position
and momentum operators~(c). The ratio of the Compton wavelength to
the oscillator length is $\lambda=10$. Position and momentum are
given in dimensionless units (see Appendix).}
\end{figure}

The Wigner function can be experimentally reconstructed in view of
the quantum tomography method~\cite{b16a2}. The problem regarding the
consistent interpretation of the quadrature operator
(or position and momentum operators) measurement arises here.
Indeed, in the relativistic case, these operators, generally
speaking, are not one-particle operators. The possibility of using the
one-particle formalism has to be clarified in this case.

Our point of view is as follows. The consistent development of the
relativistic quantum theory is possible within the framework of
the second quantization method only. However, under conditions
where the particle-pair creation does not take place, one can
consider the
 one-particle sector of the theory.
On the other hand, when one
 measures the quadrature
(position, momentum) operator, the one-particle state is
destroyed. From the viewpoint of measurement theory, the resulting
state would be an eigenstate of the operator measured. However,
This not possible because in this case one gets a state that is a
superposition of states with different
 charge signs. One can suppose that this superposition is instantly
destroyed. As a result, after a measurement we have a
multi-particle state with pairs created from vacuum.

These assumptions explain why we call the increase in coherence in
(\ref{f73}) the effective one. In fact, the coherence does not
really increase  (expression (\ref{f70}) does not contain the
$\varepsilon$-factor). The peculiarities of the process of
multi-particle operator measurement lead to such an effect.

There is the viewpoint that  strongly localized states (with
dispersion less than the Compton wavelength) in relativistic
quantum mechanics do not exist. Indeed, for spin-0 particles, it
leads to the appearance of  states with negative
dispersions~\cite{b1}. However, in addition to the position
dispersion, in real physical systems there exists an extra
parameter (the characteristic length). For a relativistic rotator,
it is the oscillator length (see Appendix), for a free particle it
is $\hbar/\Delta p$, where $\Delta p$ is
 the
momentum dispersion.

Generally speaking, there exists one more physical reason for the
lower bound of the localization. This reason itself is related to
the definition of the relativistic position operator, and follows
from the inequality $\Delta v < c$; an analysis of this problem is
given in \cite{b16a3}. It turned out, that it is possible to
consider arbitrary localized states for particles with the energy
$E\gg mc^2$.

\begin{figure}
\begin{center}
\mbox{\includegraphics[width=0.46\textwidth,
height=0.438\textwidth, clip=]{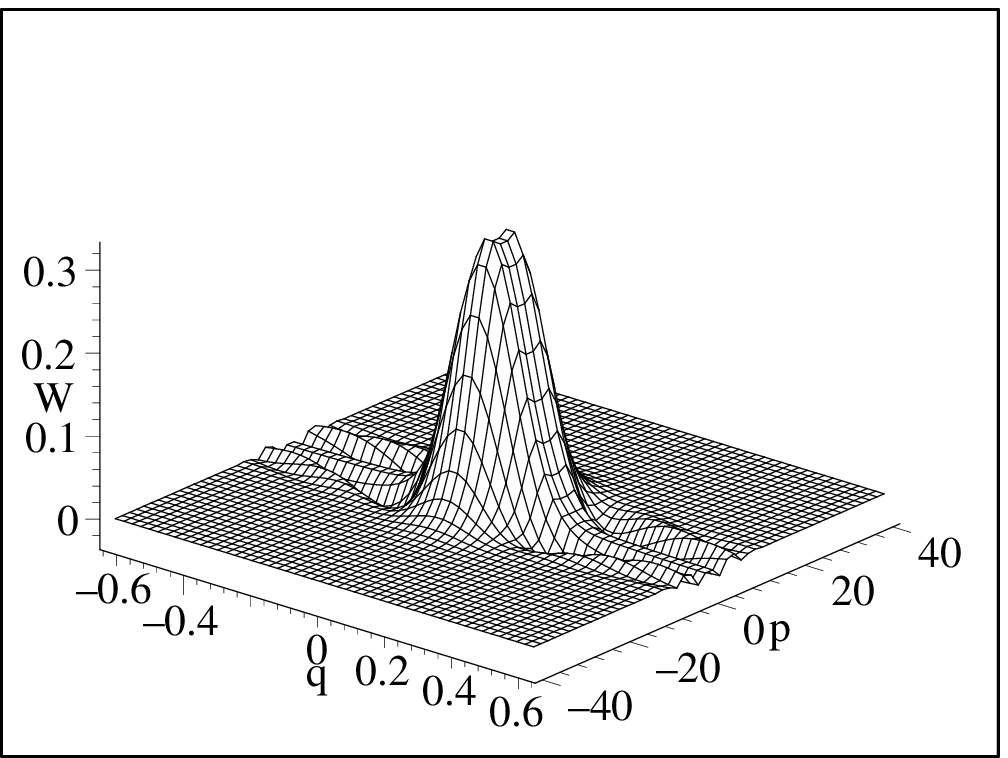}{a}
\includegraphics[width=0.46\textwidth, clip=]{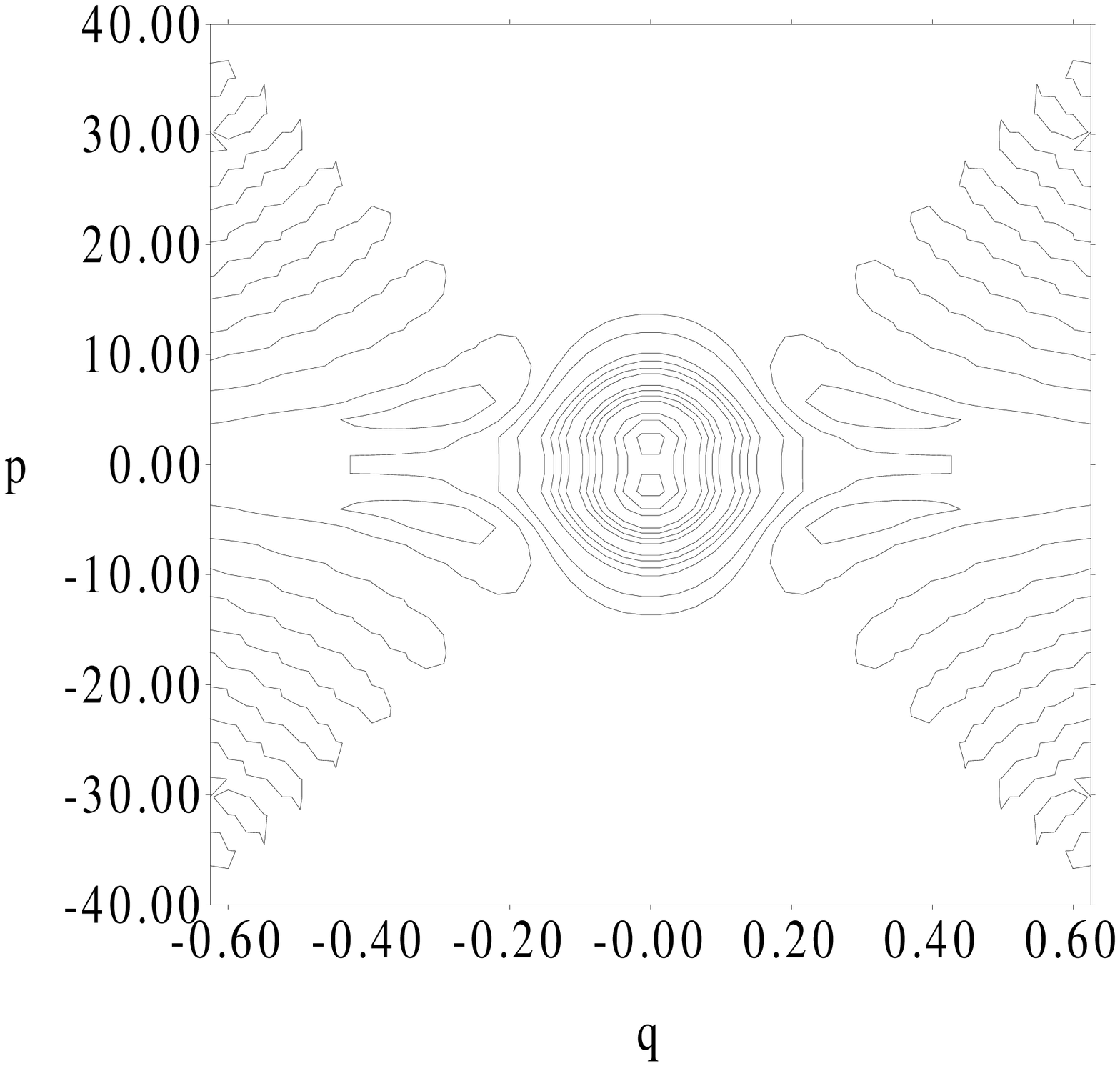}{b}}
\end{center}
\caption{\label{fig3} The Wigner function~(a) and its contours~(b)
for the Gaussin state of a free particle. The ratio
 of the Compton wavelength to the characteristic wave packet
width is $\lambda=8$. Position is given in $\lambda_c=\hbar/mc$
units and momentum, in $mc$ units.}
\end{figure}

In figure \ref{fig3} we plot the Wigner function of a free
particle for the state with the Gaussian distribution in the
momentum space. The characteristic length of this state
is 8 times less than
 the Compton wavelength.
From this figure, one can see that the particle is localized in
the $\lambda_c/8$ domain (of the order of the characteristic
length). The vacuum processes lead to the appearance of  specific
perturbations, which gives a negative contribution into
dispersion.

Another example is the eigenstate of a relativistic rotator. It
can have an arbitrary small dispersion without any perturbation.
This fact can be explained by Property \ref{p5} of the previous
Section.

An additional difficulty in the experimental observation of these
peculiarities is in the fact that they can appear in nonstationary
process only. It is well known that interference terms oscillate
with frequency
\begin{equation}
\omega_{mn}=\frac{E(m)-E(n)}{\hbar}. \label{f74}
\end{equation}
The $\varepsilon$-factor is important in the case where the
difference between energy levels is close to $mc^2$. Hence, to
verify these peculiarities, one needs to control time intervals
smaller than the Compton time
\begin{equation}
t_c=\frac{\hbar}{mc^2}. \label{f75}
\end{equation}
For $\pi^\pm$ mesons, this time equals $4.7\cdot 10^{-24}$~s, for
electrons, $1.3\cdot 10^{-21}$~s.

Similar peculiarities can appear in other systems
with the band structure of energy spectrum. This can take place,
for example,
 in semiconductors, where the analog of the Compton
time is close to $10^{-15}$~s. The nontrivial charge structure of
the position (momentum) operator, in this language, means that an
 eigenfunction of this operator is a superposition
of the states of the
 conduction and valence bands.
Measuring such an observable for conduction electrons results in a
displacement of the electrons from the valence band to the
conduction band and the appearance of electron--hole pairs.

\section{Conclusions}
\label{s7}

In this work, we have considered some mathematical peculiarities
of the phase space representation for spin-0 particles and their
physical consequences. We have restricted ourselves by such a
class of observables, whose matrix-valued Weyl symbols are
proportional to the identity matrix. We call them charge-invariant
observables. In  fact, any combination of the position and
momentum belongs to this class. However, such observables as
energy and current are not charge-invariant ones due to the
nontrivial dependence on the charge variable.

The time evolution in the standard theory is the same as in
nonlocal theory, i.e., it does not depend on the nontrivial charge
structure of position and momentum operators. However, in both
cases the time evolution differs of the evolution of classical
systems. The reason consists not only in the
 definition of the Moyal bracket and star-product.
The classical
 Hamilton function does not coincide with symbol
that plays the role of the Hamiltonian in the  evolution
equations. In the quantum
 case, the square root is determined by means of
the star-product, and relativistic effects result both in a
 large value of the nonrelativistic Hamilton
function (large
 momentum for the free particle case) and small
characteristic length of
 the system.

The nontrivial charge structure of the position  (momentum)
operators leads to the peculiarities of the constraint on the
Wigner function. However, they can appear in nonstationary
processes only, including those that are described in
nonequilibrium statistical physics. This can be explained by the
peculiarities of the relativistic position (momentum) measurement.
The initial one-particle state is destroyed in these processes. As
a result, one obtains a multi-particle state. This leads to the
appearance of additional multipliers for the interference terms
between eigenstates of the Hamiltonian in the distribution
function. These terms are responsible for the nonstationary
processes.

It is very important that these multipliers ($\varepsilon$-factor)
exceed unity. This means an effective increase in coherence in
such systems (or, to be more precise, in such kinds of
measurements). To verify this in
 the
experiment, one needs to control time intervals close to the
Compton time. However, such peculiarities can appear in other
systems with the band structure of the energy spectrum.
 For example, one
can use semiconductors where the analog of the Compton time is
close to $10^{-15}$~s. We hope that such kinds of  experiments are
possible using modern devices.

\section*{Appendix. Relativistic rotator}

Consider a particle in a constant homogeneous magnetic field and
choose the vector potential in the form
\begin{equation}
\vec{A}=\frac{1}{2}\left[\vec{B}\times\vec{r}\right], \label{af1}
\end{equation}
where $\vec{B}$ has only $z$ component
\begin{equation}
\vec{B}=\left(\begin{array}{c}0\\0\\B\end{array}\right)\label{af2}
\end{equation}
In this case, the Hamiltonian (\ref{f3}) can be written as follows
\cite{b16b}:
\begin{equation}
\hat{H}=\left(\tau_3+i\tau_2\right)\left(\frac{\hat{p}_z^2}{2m}+
\frac{\hbar\omega_c}{2}\left(\hat{\pi}^2+\hat{\xi}^2\right)\right)
+\tau_3mc^2, \label{af3}
\end{equation}
where $\omega_c=eB/m$ is the cyclotron frequency and $\hat{\pi}$ and
$\hat{\xi}$ are dimensionless linear combinations of
 momentum and position. Their physical meaning consists in describing
  the particle position in a reference frame connected with the center
  of cyclotron motion.

In our consideration, we neglect the translation motion along
the $z$ axis and consider the Hamiltonian of a relativistic rotator
\begin{equation}
\hat{H}=\frac{\hbar\omega_c}{2}\left(\tau_3+i\tau_2\right)
\left(\hat{p}^2+\hat{q}^2\right) +\tau_3mc^2. \label{af4}
\end{equation}
In the nonlocal theory representation, this Hamiltonian has a
simple form
\begin{equation}
\hat{H}^{nl}=\tau_3mc^2\sqrt{1+\lambda^2\left(\hat{p}^2+\hat{q}^2\right)},
 \label{af5}
\end{equation}
where $\lambda^2=\hbar\omega_c/mc^2$. In other words, if one
considers the characteristic oscillator length
$\sigma^2=\hbar/m\omega_c$, $\lambda$ is the ratio of the Compton
wavelength and the characteristic oscillator length.

The energy spectrum for this problem, in agreement with
(\ref{f6}), is expressed through the harmonic
oscillator spectrum and can be written in the form
\begin{equation}
E(n)=mc^2\sqrt{1+2\lambda^2\left(n+\frac{1}{2}\right)}
.\label{af6}
\end{equation}

The matrix of the Hermitian generalization of the Wigner function
for the relativistic rotator coincides with that used for the
usual harmonic oscillator and is written in the following form
\cite{b12}: \setlength\arraycolsep{2pt}
\begin{eqnarray}
&W_{nm}&(p,q)=\frac{1}{\pi}\exp\left(-q^2-p^2\right)\nonumber\\
&\times&\left\{\begin{array}{ll}
\left(\sqrt{2}\left(q-ip\right)\right)^{n-m}(-1)^m\sqrt{\frac{m!}{n!}}
L^{n-m}_m\left(2\left(p^2+q^2\right)\right) & \textrm{if $n\geq
m$}\\
\left(\sqrt{2}\left(q+ip\right)\right)^{m-n}(-1)^n\sqrt{\frac{n!}{m!}}
L^{m-n}_n\left(2\left(p^2+q^2\right)\right) & \textrm{if $n\leq
m$}
\end{array}\right., \label{af7}
\end{eqnarray}
where $L^k_l(x)$ is the generalized Laguerre polynomial. For
diagonal elements, it reads
\begin{equation}
W_{nn}(p,q)=\frac{1}{\pi}\exp\left(-q^2-p^2\right)(-1)^n
L_n\left(2\left(p^2+q^2\right)\right). \label{af8}
\end{equation}

Let us calculate the Hamiltonian (given for this problem by
(\ref{f42})). To do this, we use the fact that the Weyl symbol of
an arbitrary operator $\hat{A}$ can be expressed through the
matrix elements as follows:
\begin{equation}
A(p,q)=(2\pi)^d\sum_{n,m=0}^{\infty}A_{nm}W_{mn}(p,q).\label{af9}
\end{equation}
In our case, this expression can be written in the form
\setlength\arraycolsep{2pt}
\begin{eqnarray}
&E&(p,q)=2\pi\sum_{n=0}^{\infty}E(n)W_{nn}(p,q)\nonumber\\
&=&2mc^2e^{-q^2-p^2}\sum_{n=0}^{\infty}
\sqrt{1+2\lambda^2\left(n+\frac{1}{2}\right)}(-1)^n
L_n\left(2\left(p^2+q^2\right)\right) .\label{af10}
\end{eqnarray}

Consider the function
\begin{equation}
F(x,z)=\sum_{n=0}^{\infty}
\sqrt{1+2\lambda^2\left(n+\frac{1}{2}\right)}
L_n\left(x\right)z^n.\label{af11}
\end{equation}
Using the fact that
\begin{equation}
n^kz^n=(z\partial_z)^kz^n, \label{af12}
\end{equation}
in view of the identity for generating function for the Laguerre
polynomials \cite{b17}
\begin{equation}
\sum_{n=0}^{\infty}L_n(x)z^n=
\left(1-z\right)^{-1}\exp\left(\frac{zx}{z-1}\right)\label{af13}
\end{equation}
one can write the function (\ref{af11}) in the following form:
\begin{equation}
F(x,z)=\sqrt{1+2\lambda^2\left(z\partial_z+\frac{1}{2}\right)}
\left(1-z\right)^{-1}\exp\left(\frac{zx}{z-1}\right).\label{af14}
\end{equation}

The Hamiltonian (\ref{af10}) in our representation can be written
as follows:
\begin{equation}
E(p,q)=2mc^2\exp\left(-q^2-p^2\right)
F\left(2\left(q^2+p^2\right),-1\right). \label{af15}
\end{equation}

The term under the square root in (\ref{af14}) can
 be
expanded in a power series and one can obtain the Hamiltonian with
relativistic corrections;  we will write it here up to the
third-order terms \setlength\arraycolsep{0pt}
\begin{eqnarray}
E(p,q)=mc^2+&&\hbar\omega_c\left\{\frac{\lambda^2}{8}+
\frac{1}{2}\left(p^2+q^2\right)\left(1-\frac{5\lambda^4}{8}\right)\right.
\nonumber \\ &&\left.
-\frac{\lambda^2}{8}\left(p^2+q^2\right)^2+\frac{\lambda^4}{16}
\left(p^2+q^2\right)^3+\ldots\right\}.\label{af16}
\end{eqnarray}

This expression differs from a similar expansion for the classical
Hamilton function. Both the usual classical expression
 $\left(p^2+q^2\right)$ and $\lambda^2$ are independent relativistic
parameters of the expansion.

\end{document}